\documentclass[12pt]{article}
\usepackage{latexsym}

\newcommand\be{\begin{equation}}
\newcommand\ee{\end{equation}}
\newcommand\ba{\begin{eqnarray}}
\newcommand\ea{\end{eqnarray}}
\newcommand\str{{\rm str}}

\begin{document}

\begin{center}
\vspace{10mm}
{\large\bf Double poles in Lattice QCD with mixed actions}
\\[12mm]
Maarten Golterman,$^a$\ \ Taku Izubuchi$^b$\ \ and \ \ Yigal Shamir$^c$
\\[8mm]
{\small\it
$^a$Department of Physics and Astronomy,
San Francisco State University\\
San Francisco, CA 94132, USA}\\
{\tt maarten@stars.sfsu.edu}
\\[5mm]
{\small\it $^b$RIKEN-BNL Research Center, Brookhaven National
Laboratory\\
Upton, NY 11973, USA, and\\
Institute for Theoretical Physics, Kanazawa University\\
Kakuma, Kanazawa, 920-1192, Japan}
\\
{\tt izubuchi@quark.phy.bnl.gov}
\\[5mm]
{\small\it $^c$School of Physics and Astronomy\\
Raymond and Beverly Sackler Faculty of Exact Sciences\\
Tel-Aviv University, Ramat~Aviv,~69978~ISRAEL}\\
{\tt shamir@post.tau.ac.il}
\\[10mm]
{ABSTRACT}
\\[2mm]
\end{center}

\begin{quotation}
We consider effects resulting from the use of different discretizations for the valence
and the sea quarks, considering Wilson and/or Ginsparg--Wilson fermions.  We assume
that such effects appear through scaling violations that can be studied using
effective-lagrangian techniques.  We show that a double pole is present in
flavor-neutral Goldstone meson propagators, even if the flavor non-diagonal Goldstone
mesons made out of valence or sea quark have equal masses.  We then 
consider some observables known to be anomalously sensitive to the
presence of a double pole.  We find that the double-pole enhanced scaling
violations may turn out to be rather small in practice.
\end{quotation}

\newpage

\section{Introduction}

The use of mixed actions in Lattice QCD (LQCD) is at present gaining popularity.
With ``mixed action" we refer to any simulation in which different lattice
fermions are used in the valence and sea sectors.  (If only the bare masses
differ, we are dealing with a partially-quenched (PQ) theory \cite{bgpq}.)
The main reason is that the generation of dynamical-fermion configurations
is very expensive, while for many quantities the ``good quality" of symmetries
(such as chiral and flavor or taste symmetries) is more important in the
valence sector than in the sea sector.  An example of this are the power-like
subtractions needed for $K\to\pi$, for which good chiral symmetry in the
valence sector is essential, but not that in the sea sector.

However, the use of mixed actions raises field-theoretical questions.
At non-vanishing lattice spacing $a$ a mixed-action theory is not 
unitary, and the question is what happens to unitarity in the continuum
limit, as well as how big the effects of the violation of unitarity are at
the lattice spacings used in actual simulations.  We note that similar
issues arise for many discretizations of LQCD which do not employ 
mixed actions, such as improved actions and actions with 
Ginsparg--Wilson (GW) fermions.

Here we address the second question, the size of effects at non-zero
lattice spacing.   We extend the notion of universality to assume that
unphysical effects due to the use of a mixed action vanish in the
continuum limit, and are controlled by positive powers of $a$.  
We further assume that effective-field theory (EFT) techniques
can be used to investigate the behavior of mixed-action theories
at low energy.

The most salient phenomenon which occurs if the valence quarks
don't match the sea quarks is the appearance of a double pole in
flavor-neutral Goldstone meson propagators, with a residue $R$
proportional to $a^2$, as will be explained below.  If also the
renormalized quark masses are not equal, there is also a 
contribution from the quark mass difference, $R\propto
c_1 a^2+c_2(m_{sea}-m_{valence})$ \cite{bgpq}.  Since the
double pole represents a dramatic change of the infrared (IR) behavior of
the theory, it is important to investigate the most serious
consequences of the double pole.  As an example of this, we
consider the enhanced finite-volume effects in $I=0$ pion scattering.
For a more detailed account of this work, see Ref.~\cite{gis}.
In our work, we deal only with the cases in which the fermions in
both sectors are Wilson-like or GW fermions.\footnote{For
mixed actions with staggered sea, see Ref.~\cite{bbrs}.}

\section{The double pole}

According to our assumptions, a mixed-action theory is a PQ theory
in the continuum limit \cite{gis}.  We therefore
start with continuum chiral perturbation theory (ChPT) at lowest
order, described by the PQ lagrangian \cite{bgpq}
\be
L_{cont}=\frac{1}{8}f^2\;\str(\partial_\mu\Sigma\partial_\mu\Sigma^\dagger)
-\frac{1}{4}f^2B_0\;\str(\Sigma M^\dagger+M\Sigma^\dagger)\ ,
\ee
where $\Sigma={\rm exp}(2i\Phi/f)$ is the non-linear meson field, $f$ and
$B_0$ are the well-known lowest-order low-energy constants (LECs), and
\tolerance=2000
$M={\rm diag}(m_v,m_v,\dots,
m_s,m_s,\dots,m_v, m_v,\dots)$
the
quark mass matrix, with $K$ valence quarks of mass $m_v$, $N$
sea quarks of mass $m_s$, and $K$ ghost quarks of mass $m_v$.%
\footnote{We take degenerate valence and degenerate sea quarks,
but it is straightforward to generalize.}  The relevant chiral symmetry
group is thus $SU(K+N|K)_L\times SU(K+N|K)_R$.

\tolerance=1000
The next step we need is provided by the systematic analysis of Ref.~\cite{brs},
where mixed actions with a Wilson sea and GW valence quarks were
considered in ChPT up to order $a^2$.  It is very easy to generalize that
work to the slightly more general case where the valence and sea sectors
can be any combination of Wilson-like and GW fermions.\footnote{With
``Wilson-like" we refer to any type of Wilson fermion, including for instance
clover and twisted-mass Wilson fermions.  Our conclusions can be
straightforwardly generalized to these situations.  With GW fermions we
refer to overlap fermions or domain-wall fermions with a small enough
residual mass.  A potentially interesting special case is the combination
of untwisted Wilson sea quarks and twisted Wilson valence quarks.
If this is done by only twisting the mass matrix of the valence quarks,
this is a PQ theory, and not a (genuinely) mixed theory.  The relevant
symmetry group in that case is $SU(K+N|K)$.}

The first observation is that at $a\ne 0$ the symmetry group is smaller.
For GW valence quarks and a Wilson-like sea the symmetry group is
\be
\label{GWWsymm}
[SU(K|K)_L\times SU(K|K)_R]\times SU(N)\ \ \ \ \ ({\rm GW-Wilson})\ ,
\ee
while the case of Wilson-like valence and sea quarks (but with 
unequal lattice actions) the symmetry group is even smaller:
\be
\label{WWsymm}
SU(K|K)\times SU(N)\ \ \ \ \ ({\rm Wilson-Wilson})\ ,
\ee
because all chiral symmetries are now broken.

Up to order $a^2$ there are a number of new operators which appear
in the effective lagrangian, and we only give the ones here which are
relevant for our investigation,
\ba
\label{deltaL}
\delta L_W&=&-\frac{(af)^2}{32}\Bigl(\gamma_{vv}(\str(P_v(\Sigma-\Sigma^\dagger)))^2
+\gamma_{ss}(\str(P_s(\Sigma-\Sigma^\dagger)))^2\\
&&\hspace{2cm}
+\gamma_{vs}\str(P_v(\Sigma-\Sigma^\dagger))\str(P_s(\Sigma-\Sigma^\dagger))
\Bigr)\ ,\nonumber
\ea
in which $\gamma_{vv,vs,ss}$ are new LECs, and $P_{v,s}$ are projectors
on the valence and sea sectors, respectively.  We note that $\gamma_{vv}
=\gamma_{vs}=0$ when the valence quarks are GW, because these operators
break chiral symmetry, and thus are excluded by the symmetry group
(\ref{GWWsymm}), but not by the group (\ref{WWsymm}).

In order to obtain the meson propagators, we need to expand the lagrangian
to quadratic order in the field $\Phi$.  If we integrate out the sea-$\eta'$
(which is heavy because of the singlet part of the $\eta'$ mass), we have that
$\str(\Phi)=\str((P_v+P_s)\Phi)=0$, and Eq.~(\ref{deltaL}) reduces to quadratic order to
\be
\delta L_W=\frac{1}{2}a^2\left(\gamma_{vv}+\gamma_{ss}-2\gamma_{vs}\right)
(\str(P_v\Phi))^2+\dots\ .
\ee
Recognizing that $\str(P_v\Phi)$ is nothing else than the ``super-$\eta'$" field of 
quenched QCD, it is obvious that $\delta L_W$ leads to a double pole in
flavor-neutral propagators.

In the flavor non-diagonal sector, we find (tree-level) Goldstone meson masses \cite{brs}
\ba
M_{vv}^2&=&2B_{0v}m_v+2W_{0v}a+2\beta_v a^2\ ,\\
M_{ss}^2&=&2B_{0s}m_s+2W_{0s}a+2\beta_s a^2\ ,\nonumber
\ea
where we now allowed for the fact that with the reduced symmetry groups the
LECs $B_0$, $W_0$ and $\beta$ can be different in the valence and sea sectors.
These equations tell us (to leading order in ChPT) what it means to set
renormalized valence and sea quark masses equal: one chooses $m_{v,s}$ 
such that $M_{vv}=M_{ss}$.  We note that $W_{0v,s}=0$ if order-$a$ improved
Wilson fermions are used, and that $W_{0v,s}=0$ as well as $\beta_{v,s}=0$
for GW fermions.

Flavor-neutral propagators are given by ($i$ and $j$ are flavor indices)
\be
\langle\Phi_{ii}\Phi_{jj}\rangle(p)=
\left(\delta_{ij}-\frac{1}{N}\right)\frac{1}{p^2+M_{vv}^2}
-\frac{R}{(p^2+M_{vv}^2)^2}\ ,
\ee
with
\be
R=\frac{1}{N}\left(M_{ss}^2-M_{vv}^2\right)+
\left(\gamma_{vv}+\gamma_{ss}-2\gamma_{vs}\right)a^2\ .
\ee
It is clear that $R\ne 0$ even if $M_{vv}=M_{ss}$.  In fact, one may either
choose the valence quark mass such that $R=0$, or choose it such that 
$M_{vv}=M_{ss}$
and live with a non-vanishing $R$.  Either way, the appearance of a
double pole due to scaling violations in mixed actions is relevant for
hadronic quantities sensitive to the double pole, especially if the effects
are enhanced because of the IR-singular nature of the double pole.

\section{Example: $I=0$ pion scattering}

It is easy to find quantities particularly sensitive to the double
pole.  Examples are the $I=0$ two-pion energy in a finite volume
(which are related to the corresponding phase shifts) \cite{bg2pi},
the anomalous large-time behavior of the $a_0$ propagator
\cite{bardeenetal}, and the anomalous large-distance behavior
of the nucleon-nucleon potential \cite{bs}.  Here we will just 
briefly consider the finite-volume two-pion energy; for the $a_0$
we refer to Ref.~\cite{gis}.

The energy of two pions at rest in an $I=0$ state, and in a finite spatial
box of linear dimension $L$ with periodic boundary conditions,
is given by (to one loop) \cite{bg2pi,gis}
\be
\label{2pienergy}
\frac{\Delta E_{I=0}}{2M_{vv}}=
-\frac{7\pi}{8f^2M_{vv}L^3}+\frac{1}{2}B_0(M_{vv}L)\delta^2
+\frac{1}{2}\left(1-\frac{1}{N}\right)A_0(M_{vv}L)\delta\epsilon
+O(\epsilon^2)\ ,
\ee
where
\ba
\delta&=&\frac{R}{8\pi^2 f^2}\ ,\ \ \ \ \ \epsilon=\frac{M_{vv}^2}{16\pi^2 f^2}\ ,\\
B_0(ML)&=&-0.53+O\left(1/(ML)^2\right)\ ,\nonumber\\
A_0(ML)&=&49.59/(ML)^2+O\left(1/(ML)^3\right)\ .\nonumber
\ea
Setting $M_{vv}=M_{ss}\equiv M$, we may consider two different
regimes, one in which $\epsilon\sim M^2/\Lambda^2\sim a\Lambda_{QCD}$
\cite{brs}, and one in which $\epsilon\sim M^2/\Lambda^2\sim (a\Lambda_{QCD})^2$
\cite{aoki}, where $\Lambda\sim 1$~GeV is the chiral symmetry breaking scale.
In the first of these two regimes, the ratio of the one-loop to the tree-level
terms in Eq.~(\ref{2pienergy}) is of order $\epsilon^3\times(ML)^3$ and
$\epsilon^2\times ML$ for the $\delta^2$, respectively, $\delta\epsilon$ terms.
In the second regime this ratio is of order $\epsilon\times
(ML)^3$.

One notes that the one-loop contributions are suppressed by positive powers
of $\epsilon$ in both regimes, but for the positive powers of $ML$ which accompany
these powers of $\epsilon$.  This is an example of the enhancement due to the
``sick" IR behavior of the double pole --- these terms would all disappear when
$R=0$.  But in a mixed-action theory at $a\ne 0$, $R\ne 0$ even at $M_{vv}=M_{ss}$.

To get an idea about the size of these effects, one may substitute some typical
values for the parameters of a LQCD simulation.  Choosing for example
$a\Lambda_{QCD}\sim 0.1$, $aM\sim 0.2$ and $L/a\sim 32$, one finds 
that the $\delta$ terms in Eq.~(\ref{2pienergy}) represent scaling violations
of order 10\%.  This is small, but not negligible.  Of course, in this estimate,
we took all other constants (including the coefficient of $a^2$ in $R$) equal
to one.  Actual simulations are needed to make further progress with the
investigation of the size of these effects.

\section{Conclusion}

Let us summarize what we learned about the use of mixed actions for
LQCD computations.  

We begin by re-emphasizing that we assumed that the unphysical effects
of mixed actions are encoded in scaling violations, and that EFT
techniques can be used to study the issue.  We recall that this approach
has proven very successful in similar cases, such as the unphysical
effects due to (partial) quenching.  It is therefore important to test this
assumption also in this case through actual simulations, in which in addition
the numerical size of the effects can be estimated more reliably than through our
parametric estimates above and in Ref.~\cite{gis}.

We have concentrated on the role of the double pole, because it is
the most IR-sensitive probe of unphysical effects due to the use of
mixed actions, just as it is in the case of (partial) quenching.  Clearly,
the effects are quantity dependent, and are expected to be most
pronounced for those quantities for which there is a clear unphysical
enhancement.  The example we gave is that of enhanced finite-volume
effects in pion scattering.   We concluded in that case that the effects are
likely to be rather small numerically in present-day simulations, but
not so small as to be automatically negligible.

Finally, we observe that in the case of the $I=0$ two-pion energy,
this effect can also be monitored if one leaves out the ``double-annihilation" 
diagram (as was done in Ref.~\cite{jlqcd} because of poor
statistics), because the effective theory can be adapted to reflect
this situation \cite{bg2pi}.  For the terms shown in Eq.~(\ref{2pienergy})
this corresponds to dropping the $\delta^2$ term.  Our 10\% estimate
above came largely from the $\delta\epsilon$ term, and it is thus
plausible that the size of the effect does not become much smaller
without the double-annihilation diagram.

\section*{Acknowledgments}

We thank Oliver B\"ar and Steve Sharpe, and in particular Paulo Bedaque and Claude Bernard,
for helpful discussions.
This work was started during a visit of two of us (MG and YS) to
Brookhaven National Lab and Columbia University, and we thank the RBC collaboration
for discussions and hospitality.  MG also thanks the Lawrence Berkeley
National Lab Nuclear Theory Group for hospitality.
MG is supported in part by the US Department of Energy, and
YS is supported by the Israel Science Foundation under grant
222/02-1.

\end{document}